   \documentclass[]{article}
  \usepackage{amsmath, amsthm, amssymb,graphics,epsf,authblk,wrapfig}
  \usepackage[comma,numbers]{natbib}
  \usepackage[font=small,labelfont=bf]{caption}
  \usepackage{parskip}
  \newcommand{\f}{\frac}
  \newcommand{\Lm}{\Lambda}

\newcommand{\tnorm}[1]{\left\| #1 \right\|_2}

\newcommand{\pderiv}[3]{\f{
\partial #1}{
\partial #2}}

  \usepackage{amsmath, amsthm, amssymb,graphics,epsf,epstopdf,graphicx,geometry,mathrsfs}
  \usepackage{listings}
  \usepackage{listings}
\usepackage{color}
\usepackage{cancel}
 
\definecolor{dkgreen}{rgb}{0,0.6,0}
\definecolor{gray}{rgb}{0.5,0.5,0.5}
\definecolor{mauve}{rgb}{0.58,0,0.82}
 
\title{PrAGMATiC: a Probabilistic and Generative Model of Areas Tiling the Cortex}  
\author[1]{Alexander G. Huth\thanks{This work was supported by grants from the National Science Foundation (IIS1208203), the National Eye Institute (EY019684), and from the Center for Science of Information (CSoI), an NSF Science and Technology Center, under grant agreement CCF-0939370. AGH was also supported by the William Orr Dingwall Neurolinguistics Fellowship. We thank Wendy de Heer, Jascha Sohl-Dickstein, Keenan Crane, and Natalia Bilenko for useful technical discussions.}}
\author[1,2,3]{Thomas L. Griffiths}
\author[1,2]{Frederic E. Theunissen}
\author[1,2]{Jack L. Gallant}
\affil[1]{Helen Wills Neuroscience Institute, University of California, Berkeley}
\affil[2]{Department of Psychology, University of California, Berkeley}
\affil[3]{Cognitive Science, University of California, Berkeley}
\renewcommand\footnotemark{}

\begin{document}
\maketitle

\begin{abstract}
Much of the human cortex seems to be organized into topographic cortical maps. Yet few quantitative methods exist for characterizing these maps. To address this issue we developed a modeling framework that can reveal group-level cortical maps based on neuroimaging data. PrAGMATiC, a probabilistic and generative model of areas tiling the cortex, is a hierarchical Bayesian generative model of cortical maps. This model assumes that the cortical map in each individual subject is a sample from a single underlying probability distribution. Learning the parameters of this distribution reveals the properties of a cortical map that are common across a group of subjects while avoiding the potentially lossy step of co-registering each subject into a group anatomical space. In this report we give a mathematical description of PrAGMATiC, describe approximations that make it practical to use, show preliminary results from its application to a real dataset, and describe a number of possible future extensions.
\end{abstract}

In the best understood regions of the human neocortex, functional representations appear to be organized into topographic cortical maps: retinotopic and semantic maps in visual cortex \citep{Sereno1995,Huth2012}, tonotopic maps in auditory cortex \citep{Romani1982}, somatotopic maps in somatomotor cortex \citep{Penfield1937}, and numerosity maps in parietal cortex \citep{Harvey2013}. It seems likely that other regions of the brain--such as prefrontal cortex--are also organized into cortical maps, but those maps have yet to be discovered. One major obstacle for defining cortical maps is that both anatomy and functional representation can differ substantially between individuals, thwarting current methods for combining data across subjects \citep{Amunts2000,Crivello2002,Fedorenko2010,Haxby2011}. 

The primary question in combining data across subjects is that of correspondence: how do we determine which response in subject A corresponds to a given response in subject B? One possibility is by using anatomy. Many existing methods align anatomical images of brains from multiple subjects in either volumetric space \citep{Jenkinson2001} or surface space \citep{Fischl1999}. These methods assume that brain anatomy and function are perfectly correlated, and that any deviation from this relationship is noise. Yet this assumption has repeatedly been proven false \citep{Crivello2002,Fedorenko2010}. Indeed, even low-level functional areas such as V1 can vary significantly in size across individuals \citep{Amunts2000}.

Another way to find correspondence between subjects is by using functional responses. Some new methods find corresponding temporal patterns of functional activation in each subject given responses to a complex natural stimulus, such as a movie \citep{Haxby2011,Bilenko}. These methods have proven much more successful at predicting functional data in new subjects than anatomical methods (and may in fact be near-optimal for mapping data from one subject to another). Working purely in functional space makes it possible to disregard anatomical differences between subjects, but it also means that these methods provide little information about the organization of cortical maps (indeed they would work just as well even if cortical maps were completely different in each subject). We might be able to learn more about how functional representations on the cortex are organized if we can take both function and anatomy into account.

\begin{figure}[p!]
  \begin{center}
    \includegraphics[width=0.52\textwidth]{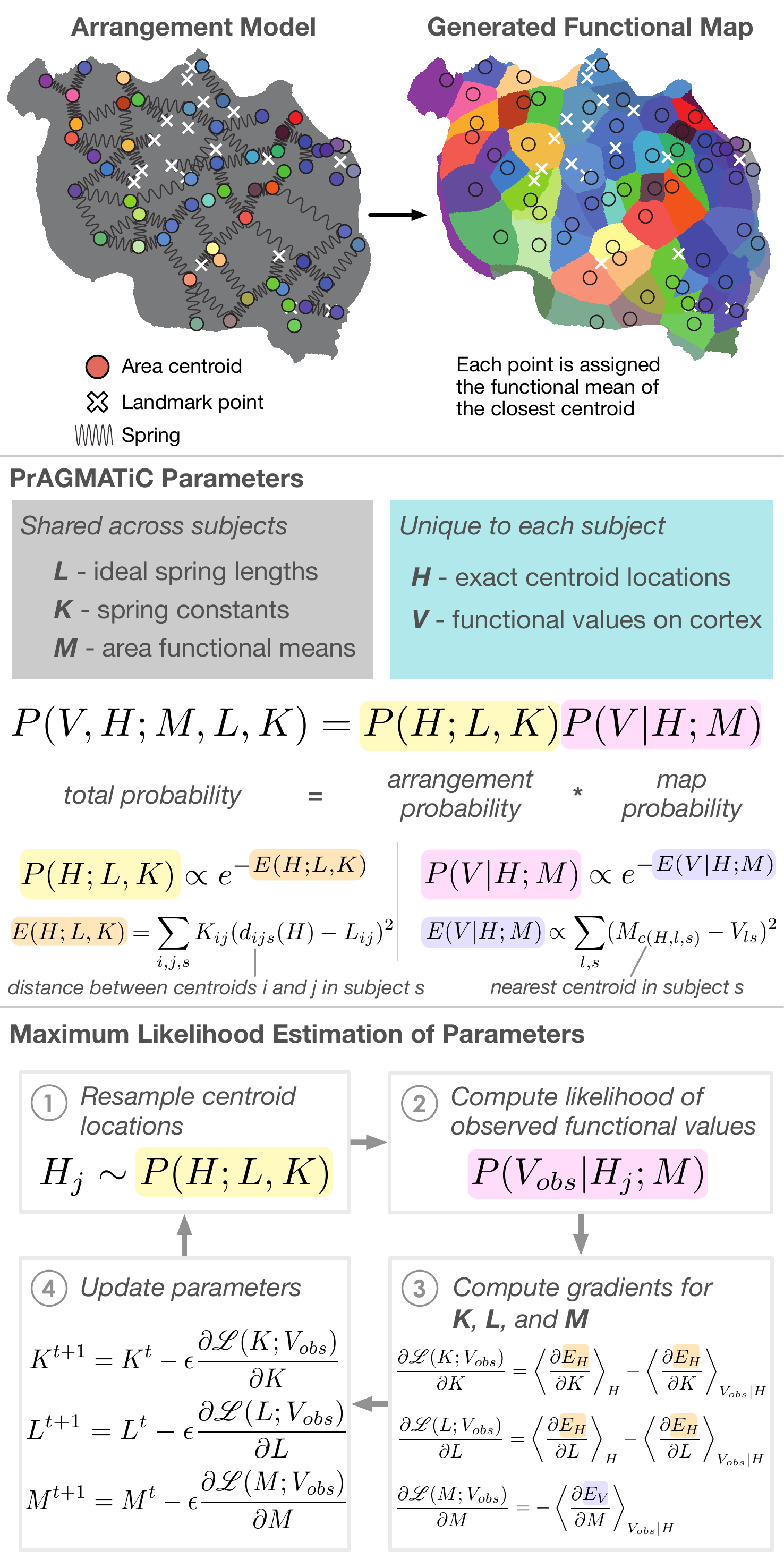}
  \end{center}
  \caption{\textbf{PrAGMATiC model schematic.} In PrAGMATiC the arrangement of areas in a cortical map is modeled as a physical system of springs, where each pair of area centroids is bound together by a spring. To generate a map, each point on the cortex is assigned the functional mean of the closest area centroid. Probability distributions over arrangements and maps are formed using Boltzmann distributions. The parameters of these distributions are constrained to be common across subjects, but the exact centroid locations are allowed to vary. Parameters are learned using maximum likelihood estimation.}
  \vspace{-50pt}
\end{figure}

A third way to find correspondence between subjects is the region-of-interest (ROI) analysis \citep{Saxe2006}, which combines both anatomical and functional constraints. In this method, an ROI is defined in each subject using a functional localizer contrast. Among voxels that respond significantly to the localizer a single cluster is selected as the ROI based on its approximate anatomical location. Then on a different dataset, average functional responses within that ROI are compared across subjects. This method does not assume exact anatomical correspondence across subjects. However, for many areas of the brain there is no known localizer, making it impossible to apply the ROI method. 

With the goal of combining the strengths of purely functional and ROI-based methods we have developed PrAGMATiC: a probabilistic and generative model of areas tiling the cortex. This hierarchical Bayesian modeling framework accounts for individual differences both in anatomy and function by treating the cortical map observed in each subject as a sample from a single underlying probability distribution. This distribution consists of two parts: an arrangement model and an emission model. The arrangement model uses a novel spring-based approach to describe cortical topography. This flexible approach can account for substantial individual differences in the shape, size, and anatomical location of functional brain areas. The emission model then generates predicted functional cortical maps based on the map arrangement.

In this paper we describe the mathematical formulation of this model and show how the various parameters can be learned using a Markov chain Monte Carlo technique similar to contrastive divergence \citep{Hinton2002}. We then show some results from applying PrAGMATiC to a simple example dataset, and describe some possible future extensions to the model.

\section{Model formulation}
In the basic instantiation of PrAGMATiC we suppose that the cortex of each subject is tiled with convex areas, and that all locations within each area have the same functional properties. This model is shown schematically in Figure 1. The location of each area is determined by the location of its centroid, which is a single point on the cortical surface. Centroid locations depend on the locations of neighboring centroids as well as the locations of some known cortical landmarks, which are identified separately in each subject. The functional properties of each area are represented as a vector. The values in this vector could be raw BOLD responses, weights from a GLM or voxelwise model, or any derived measure, such as projections of voxelwise model weights onto a set of orthogonal components \citep{Huth2012}. The mean functional value for area $i$ is denoted $M_i$.

The model for each subject is instantiated as a two-layer Bayesian network, with one visible layer and one hidden layer. The visible layer units are vertices on the cortical surface mesh. Each vertex is associated with a $D$-vector of observed functional values. The vector of observed values for visible unit $l$ in subject $s$ is called $v^{obs}_{ls} \in \mathcal{R}^D$, and the collection of all visible units in a subject is called $V^{obs}_s$. We assume that all visible units are independent of each other given the hidden layer units.

The hidden layer units are the locations of the area centroids. The location of centroid $i$ in subject $s$ is called $h_{is}$, and the collection of all hidden units in subject $s$ is called $H_s$.

As a generative model, PrAGMATiC must be able to generate samples from the distribution of visible unit vectors. To sample $v_{ls}$ we first find the index of the nearest area centroid on the cortical surface, $c(H,l,s)$. Then we look up the mean associated with that centroid, $M_{c(H,l,s)}$. Finally we draw a sample from a multivariate Gaussian distribution with spherical variance: \begin{equation}v_{ls} \sim \mathcal{N}\left( M_{c(H,l,s)}, \sigma^2_V I_D \right)\end{equation}

The probability distribution over locations of the hidden units is modeled using a physical analogy to a system of springs. The ideal length of the spring connecting units $i$ and $j$ is called $L_{ij}$, and the spring constant is $K_{ij}$.

The full probability distribution for PrAGMATiC is written as the product of the probability of the arrangement, $P(H;L,K)$, and the probability of the visible map given the arrangement, $P(V|H;M)$. Here we write these two probability distributions as Boltzmann distributions with energy functions $E(H;L,K)$ and $E(V|H;M)$.
\begin{eqnarray}
P(V,H; M,L,K) &=& P(H;L,K) P(V|H;M) \\
&=& \left[\f{e^{-E(H;L,K)}}{Z_H(L,K) }\right] \left[\f{e^{-E(V|H;M)}}{ Z_V } \right]
\end{eqnarray}
where $Z_H(L,K) = \sum_h e^{-E(h;L,K)}$ and $Z_V = \sum_v e^{-E(v|H;M)}$ are normalizing constants.

The distribution over arrangements of the hidden layer units is modeled using a Boltzmann distribution with the following energy function:
\begin{equation}E(H;L,K) = \f{\beta}{2} \sum_{i,j,s} K_{ij} \left( d_{ijs}(H) - L_{ij}\right) ^ 2\end{equation}
Here $K_{ij}$ is the spring constant for the spring linking centroids $i$ and $j$, $L_{ij}$ is the ideal length of the spring linking $i$ and $j$, and $d_{ijs}(H)$ is the geodesic distance across the cortex between centroids $i$ and $j$ in subject $s$ and in the arrangement $H$. We compute these geodesic distances using a heat-based approximation to the exact geodesic distance across the cortical surface mesh \citep{Crane2013}. This energy function, $E(H;L,K)$, is exactly the sum over the spring potential energy for all spring connections in the model. The constant $\beta$ is the inverse temperature of the spring system: if $\beta$ is large, then the system has low temperature, and the springs will not deviate much from their ideal lengths; if $\beta$ is small, then the system has high temperature, and the springs will deviate a lot from their ideal lengths. The normalizing constant for $P(H;L,K)$ depends on both the values of $L$ and $K$, and is written here as $Z_H(L,K)$.

The distribution over visible unit values is multivariate Gaussian with equal variance in all dimensions and zero covariance, but for consistency we write it as an energy-based model. The energy function for the visible units is:
\begin{equation}E(V|H;M) = \f{\sigma_V^{-2}}{2} \sum_{l,s} \tnorm{ v_{ls} - M_{c(H,l,s)} }^2 \end{equation}
Here $v_{ls}$ is the functional value for visible unit $l$ in subject $s$, and $M_{c(H,l,s)}$ is the mean functional value for the closest hidden layer unit (by geodesic distance across the cortical surface) in the arrangement $H$ to visible layer unit $l$ in subject $s$. The constant $\sigma_V$ is the standard deviation of the gaussian, which is assumed to be equal in all dimensions. The normalizing constant for $P(V|H;M)$ depends on $\sigma_V$, but not on any of the learned parameters (because this is a Gaussian distribution its normalizing constant is known). It is written here as $Z_V$.

\subsection{Maximum likelihood estimation}
We use maximum likelihood estimation (MLE) to learn the spring lengths, $L$, the spring constants, $K$, and the area functional means, $M$, based on observed visible unit data, $V^{obs}$. Here we show the derivation of the MLE learning rules from the model formulations given above.

\paragraph{Spring lengths.} The average log likelihood of the spring lengths, $L$, given the observed data is written:
\begin{equation}\mathscr{L}(L;V^{obs}) = \f{1}{N} \sum_s \log P(V^{obs}_s; L, K, M)\end{equation}
where $N$ is the number of subjects, $s$ is an index across subjects, and the total probability of the observed data given the parameters, $P(V^{obs}_s; L,K,M)$ is equal to the expectation of the full probability distribution over $H$:
\begin{equation}
P(V^{obs}_s; L,K,M) = \sum_h P(V^{obs}_s | h; M) P(h; L,K)
\end{equation}
To fit our model, we want to increase the average likelihood by changing the spring lengths, $L$. We can do this by differentiating the likelihood function with respect to $L$, and then changing $L$ by a small amount in the direction that will increase the likelihood. This process will be repeated many times.

Differentiating the likelihood with respect to $L$, we find:
\begin{eqnarray}
\pderiv{\mathscr{L}(L;V^{obs})}{L}{} &=& \left( \pderiv{Z_H(L,K)}{L}{} \right)^{-1} Z_H(L,K) \nonumber \\
&& - \f{1}{N} \sum_s \f{\sum_h P(V^{obs}_s | h; M) P(h; L,K) \pderiv{E(h;L,K)}{L}{}}{P(V^{obs}_s; L,K,M)}
\end{eqnarray}
The first part of the gradient, which involves the normalizing constant $Z_H$, can be written as:
\begin{equation} \left( \pderiv{Z_H(L,K)}{L}{}\right)^{-1} Z_H(L,K) = \f{1}{N} \sum_{s,h} P(h; L,K) \pderiv{E(h;L,K)}{L}{}\end{equation}
or simply as the expectation of the gradient over $H$:
\begin{equation} \left( \pderiv{Z_H(L,K)}{L}{}\right)^{-1} Z_H(L,K) = \left\langle \pderiv{E(H;L,K)}{L}{} \right\rangle_H\end{equation}
Thus we note that the entire gradient could be written more simply as the Boltzmann machine learning rule from Ackley, Hinton, and Sejnowski, 1985 \citep{Ackley1985}:
\begin{equation}\pderiv{\mathscr{L}(L;V^{obs})}{L}{} = \left\langle \pderiv{E(H;L,K)}{L}{} \right\rangle_H - \left\langle \pderiv{E(H;L,K)}{L}{} \right\rangle_{H|V^{obs}}\end{equation}

However, we retain the earlier formulation in order to make an essential approximation. This gradient is impossible to compute exactly because it requires taking the expectation over all possible $H$. Even though $H$ is discrete, there are far too many possible states to reasonably check. Therefore we approximate the gradient using only a small number of samples from $P(H;L,K)$. These are obtained using Gibbs sampling, wherein the location of each hidden unit is updated sequentially according to the conditional distribution $P(h_{is} | H_{\setminus is}; L,K)$, where $H_{\setminus is}$ contains the locations of all the hidden units except for unit $i$ in subject $s$. This procedure is used to obtain $J$ samples from $P(H;L,K)$, which are denoted $\tilde{H}^j$ with $j=1...J$. 

This approximation is necessary because it is difficult to compute the expected gradient conditioned on the visible units. In most energy-based models there is an analytic expression for the conditional distribution of the hidden units given the visible units, $P(H|V)$, and thus it is easy to estimate the conditional expected gradient. Here it is very expensive to sample from the conditional distribution $P(H|V)$, but it is cheap to sample from $P(H;L,K)$.

We use the samples from $P(H;L,K)$ to approximate the expectations over $H$. This allows us to rewrite the gradient function as:
\begin{eqnarray}
\pderiv{\mathscr{L}(L;V^{obs})}{L}{} &=&
\frac{1}{NJ} \sum_{j,s}  \pderiv{E(\tilde{H}^j_s;L,K)}{L}{} - 
\frac{1}{N} \sum_{s} \f{\sum_j P(V^{obs}_s | \tilde{H}^j_s; M) \pderiv{E(\tilde{H}^j_s;L,K)}{L}{}}{\sum_j P(V^{obs}_s | \tilde{H}^j_s; M)}
\end{eqnarray}

This formula shows that the likelihood gradient for $L$ is equal to the difference between the average energy gradient across all $J$ arrangement samples (the first term), and a weighted average energy gradient (the second term) where the weights are proportional to the probability of the observed data $V^{obs}$ given the sampled $H$. Intuitively, the second term increases the probability of the observed data, while the first term decreases the probability of unobserved data (i.e. random samples from the model).

To compute the energy gradient for each arrangement sample we differentiate the energy function with respect to each element of $L$, giving:
\begin{equation}\pderiv{E(H;L,K)}{L_{ij}}{} = \beta \sum_s K_{ij} \left(L_{ij} - d_{ijs}(H) \right) \end{equation}

\paragraph{Spring constants.} Similarly, the gradient of the likelihood with respect to the spring constants, $K$ is derived as:
\begin{eqnarray}
\pderiv{\mathscr{L}(K;V^{obs})}{K}{} &=&
\frac{1}{NJ} \sum_{j,s}  \pderiv{E(\tilde{H}^j_s;L,K)}{K}{} - 
\frac{1}{N} \sum_{s} \f{\sum_j P(V^{obs}_s | \tilde{H}^j_s; M) \pderiv{E(\tilde{H}^j_s;L,K)}{K}{}}{\sum_j P(V^{obs}_s | \tilde{H}^j_s; M)}
\end{eqnarray}
And the energy gradient is:
\begin{equation}\pderiv{E(H;L,K)}{K_{ij}}{} = \f{\beta}{2} \sum_s \left( d_{ijs}(H) - L_{ij} \right)^2 \end{equation}

\paragraph{Area functional means.} The gradient for $M$ is slightly different because the normalizing constant $Z_V$ does not depend on $M$ (as the normalizing constant of a Gaussian does not depend on the mean). Thus it has a simpler expression:
\begin{eqnarray}
\pderiv{\mathscr{L}(M;V^{obs})}{M}{} &=&
- \frac{1}{N} \sum_{s} \f{\sum_j P(V^{obs}_s | \tilde{H}^j_s; M) \pderiv{E(V^{obs}|\tilde{H}^j_s;M)}{M}{}}{\sum_j P(V^{obs}_s | \tilde{H}^j_s; M)}
\end{eqnarray}
And the energy gradient for the mean of area $i$, $\partial E_V / \partial M_i$ is:
\begin{equation}
\pderiv{E(V|H;M_i)}{M_i}{} = \sigma_V^{-2} \sum_{l,s; c(H,l,s)=i} \left( V_{ls} - M_{i} \right)
\end{equation}
where the sum is taken only over the visible units $l,s$ for which the closest hidden unit in the arrangement, $c(H,l,s)$ is $i$.

\subsubsection{Learning using MLE}
To obtain high quality, independent samples of $H$ we maintain $J$ parallel Markov chains for each of the $N$ subjects. At each learning step we perform one Gibbs sweep through each of the Markov chains. That is, at step $t$ in chain $j$ and subject $s$ we draw the sample: 
\begin{equation}
\tilde{H}^{j,t}_s \sim P(H_s | \tilde{H}^{j,t-1}_s;L^t,K^t)
\end{equation}
For each of the $J$ samples we compute the energy gradients for $L$, $K$, and $M$, as well as the likelihood of the observed data $P(V^{obs}_s|\tilde{H}^{j,t}_s;M^t)$. Then we compute the average gradients (for $L$ and $K$) and the weighted average gradient according to the data likelihoods. Finally we update $L$, $K$, and $M$ by taking a small step down these gradients:
\begin{eqnarray}
 L^{t+1} &=& L^t - \epsilon_L \pderiv{\mathscr{L}(L^t;V^{obs})}{L^t}{} \\
 K^{t+1} &=& K^t - \epsilon_K \pderiv{\mathscr{L}(K^t;V^{obs})}{K^t}{} \\
 M^{t+1} &=& M^t - \epsilon_M \pderiv{\mathscr{L}(M^t;V^{obs})}{M^t}{} 
 \end{eqnarray}
 
This process is then repeated for many thousands of steps. Although each step is stochastic (due to the finite number of samples, $J$, that was used to approximate the gradients), the likelihood function will tend to increase over time.
 
\subsection{Hyperparameters}
The hyperparameters $\beta$ and $\sigma_V$ affect learning speed but do not directly affect the learned parameters (except by virtue of poor approximation). The inverse spring temperature, $\beta$, determines how stiff or loose the springs are. If $\beta$ is very high, then the springs will be very stiff, samples of the arrangement $H$ will be highly correlated across iterations, and the quality of the gradient steps will suffer. If $\beta$ is very low, then the springs will be very floppy, samples of the arrangement $H$ will be highly random, and the quality of the gradient steps will also suffer. 
 
If the standard deviation of the visible unit Gaussian distribution, $\sigma_V$ is very low, then one sample from the arrangement $H$ will always yield much higher likelihood of $V^{obs}$ than the others, and the weighted average of the gradients across samples will become the difference between the best sample and the other samples. If $\sigma_V$ is very high, then the likelihood of $V^{obs}$ will be very similar for all samples, and the weighted average of the gradients will be almost identical to the simple average across gradients, making learning slow. 
 
Note further that these hyperparameters interact with each other. If the inverse temperature $\beta$ is high, then almost all samples of the arrangement $H$ will be close to the $H$ that minimizes the total spring energy. Because the samples will be more similar, the likelihoods will also be more similar, and the weighted average of the gradients will again be similar to the simple average. The hyperparameters also interact with the number of areas in the model.
 
Rather than tuning these parameters directly, we select desired levels of entropy for the distribution over each centroid given the other centroids $P_h = P(h_{is} | H_{\setminus is}; L,K)$ and the distribution over observed visible unit values given the sampled arrangements $P_V = P(V^{obs}_s | \tilde{H}^j_s; M)$. If the average entropy of the centroid distribution $P_h$ is lower than the target value, we then lower the inverse temperature $\beta$ to make the springs more floppy, and if it is higher than the target value we raise $\beta$ to make the springs stiffer. If the entropy of the visible unit distribution $P_V$ is lower than the target value, we raise the standard deviation $\sigma_V$ to give higher probability to the less optimal states, and if it is higher than the target value we lower $\sigma_V$ to give lower probability to the less optimal states. We check the entropies at the end of each learning iteration. Then we adjust $\beta$ and $\sigma_V$ by 10\% upward or downward depending on whether the entropies are too high or too low.

High entropy keeps the model from falling into local minima, but also keeps the model from finding very high likelihood solutions. Conversely, low entropy allows the model to find high likelihood solutions, but also makes it more likely to fall into local minima. To take advantage of both low and high entropy states we use an annealing approach, where the entropy target for the centroid distribution $P_h$ is high at the beginning of learning, but then is gradually lowered throughout the learning process. This makes the Markov chain take larger, more uncorrelated steps at the beginning of learning, but smaller steps at the end.

\subsection{Initialization}
Proper initialization of the PrAGMATiC parameters and hidden variables is essential for successful model fitting. In particular, the area centroids need to be initialized in each subject so that the overall topology of the spring system is preserved. If the centroids are randomly placed in each subject it is highly likely that there will be topological errors, where the entire map or portions thereof will be mirrored in some subjects relative to other subjects. Topological errors represent large local minima in the spring energy function, and are very difficult for the sampler to overcome without running at high temperature for a long time. 

To avoid this problem we construct pseudorandom arrangements where the topology is nearly identical in every subject. First we initialize the centroids in one subject. This can be done randomly or pseudorandomly. Then we compute the geodesic distance across the cortex from each randomly-placed centroid to each landmark. To initialize centroids in other subjects we then find the centroid locations that minimize the total squared difference in centroid-landmark distance from the distances we measured in the first subject. This process guarantees that the topology of the spring system is the same in every subject.

Next we decide which springs, out of all possible centroid-centroid pairs, will be included in the model. This is done using a process similar to Delaunay triangulation. We use the centroids to construct a Voronoi tessellation of the cortex in each subject by assigning every vertex on the surface to the nearest centroid. Then we find each pair of centroids that share a border in the Voronoi tessellation. This is done by checking whether each pair of neighboring vertices in the cortical surface belong to different centroids. Because the Voronoi tessellation is dual to the Delaunay triangulation, we can then construct the Delaunay triangulation by simply connecting each pair of neighboring centroids. Only these centroid-centroid springs are then considered in the spring model. All possible landmark-centroid springs are also included.

Based on the initial centroid placements we initialize the spring lengths, $L$, to be the average spring lengths across all the subjects. The spring constants, $K$, are set to 1 for all springs that are included in the model. To initialize the area functional means, $M$, we compute the mean functional values for each area in the parcellation given by the initial centroid placements.

\subsection{Improvements to stability}
In practice the algorithm as written above converges when the numbers of areas and subjects are both small. If these numbers become large, however, (e.g. 128+ clusters and 5+ subjects) the algorithm becomes less stable. When this occurs, the model tends to prioritize minimum energy solutions over maximum probability solutions. That is, the model tries to minimize the total spring energy across all the subjects at the cost of poorly explaining the data. This often causes all the centroids to cluster together as far as possible from any known landmarks. This type of solution minimizes the energy of the spring system across subjects, but does not maximize the probability of the observed data. These effects are exacerbated when $\beta$ is high.

We believe that this problem is caused by bias in drawing samples of the arrangement $H$. When the spring temperature is low, all samples from the distribution over arrangements $P(H; L,K)$ will be very close to the minimum spring energy state (i.e. the arrangement $H^-$ that minimizes the total spring energy $E(H^-;L,K)$). If the minimum spring energy state is far from the maximum probability state (i.e. the arrangement $H^*$ that maximizes the probability of the observed data $P(V^{jobs}|H^*)$), then this could bias the gradient steps such that the likelihood of the observed data decreases over time, although the spring energy also decreases.

One solution to this problem is to increase $J$, the number of samples that are drawn for each subject at each step of the learning algorithm. However, this is expensive, as run time of the learning algorithm is linear in $J$. An alternative solution that incurs almost no additional cost is to add a small perturbation to the ideal spring lengths that is tailored to each subject. That is, we replace the global spring length parameters $L$ with $L + \Lm_s$, where $\Lm_s$ is a subject-specific length parameter for subject $s$ that is also learned using MLE. With this change, the domain of possible minimum spring energy states is much larger, and thus it is much more likely that the maximum probability state also has low or minimum spring energy. To ensure that $\Lm_s$ stays small and that the average perturbation across subjects stays close to zero, we add a quadratic regularization penalty.

With this new term, the full probability distribution is now written:
\begin{eqnarray}
P(V,H | \Lm ; M,L,K) &=& P(H | \Lm; L, K) P(V|H; M) P(\Lm) \\
&=& \left[Z_H(L, \Lm, K)^{-1} e^{-E(H; L, \Lm, K)} \right] \left[ Z_V^{-1} e^{-E(V|H;M)} \right] \left[ Z_\Lm^{-1} e^{-E(\Lm)}\right]
\end{eqnarray}
where $P(\Lm)$ is a zero-mean Gaussian prior on $\Lm$.

The spring energy function becomes:
\begin{equation}
E(H|\Lm;L,K) = \f{\beta}{2} \sum_{i,j,s} K_{ij} \left( d_{ijs}(H) - L_{ij} - \Lm_{ijs} \right) ^ 2
\end{equation}
and the energy function for $\Lm$ is:
\begin{equation}
E(\Lm) =  \f{\sigma_\Lm^{-2}}{2} \sum_{i,j,s} \Lm_{ijs}^2
\end{equation}

The spring energy gradient for $\Lm$ is very similar to that for $L$:
\begin{equation}\pderiv{E(H|\Lm;L,K)}{\Lm_{ijs}}{} = \beta K_{ij} \left(L_{ij} + \Lm_{ijs} - d_{ijs}(H) \right)
\end{equation}

And the energy gradient for the Gaussian prior on $\Lm$ is simple:
\begin{equation}
\pderiv{E(\Lm)}{\Lm_{ijs}}{} =  \sigma_\Lm^{-2} \Lm_{ijs}
\end{equation}

Because $\Lm$ incorporates a prior, it has a slightly different likelihood gradient than $L$:
\begin{equation}
\pderiv{\mathscr{L}(\Lm;V^{obs})}{\Lm}{} = \left\langle \pderiv{E(H;L,K,\Lm)}{\Lm}{} \right\rangle_H - \left\langle \pderiv{E(H;L,K,\Lm)}{\Lm}{} \right\rangle_{H|V^{obs}} - \pderiv{E(\Lm)}{\Lm}{}
\end{equation}

One issue with this solution is that it changes the meaning of the spring constants, $K$. If the total subject-specific spring lengths are exactly equal to the distances between centroids (i.e. $d_{ijs}(H)=L_{ij}+\Lm_{ijs}$ for all $i$, $j$, and $s$), then the spring constants cannot be learned because the gradient for $K$ becomes zero. It might be possible to simply fix $K$ and learn $\Lm$, then later set $K$ to be the variance of $\Lm$ over subjects. In any case, it is probably unwise to learn $K$ and $\Lm$ simultaneously.

\subsection{Improvements to efficiency}

Even with the gradient approximation, this model can be prohibitively expensive to fit. The computational complexity of the parameter estimation comes mostly from two operations: drawing samples from the distribution over arrangements $P(H)$ using Gibbs sampling, and computing the likelihood of the observed visible unit data $P(V^{obs}|H)$. Of these two steps sampling $P(H)$ is by far the more expensive. However, we can make further approximations that simplify this process significantly. 

In the Gibbs sampler we update the location of each centroid, $h_{is}$, based on the locations of the other centroids, $H_{\setminus is}$. This is done by finding the total spring energy given each possible location of the selected centroid, summing these energies to find the normalizing constant, and then sampling from the resulting multinomial distribution. That is, for each possible location $h_{is}=l$, we compute the total spring energy:
\begin{equation}
E_{is}(h_{is}=l) = \f{\beta}{2} \sum_j K_{ij} \left(d_{ijs}(h_{is}=l, H_{\setminus is}) - L_{ij}\right)^2
\end{equation}
Then to find the conditional probability $P(h_{is}=l^* | H_{\setminus is})$ we divide by the sum of the energies:
\begin{equation}
P(h_{is}=l^* | H_{\setminus is}) = \f{E_{is}(h_{is}=l^*)}{\sum_l E_{is}(h_{is}=l)}
\end{equation}
To compute this distribution exactly we would need to find the total spring energy for every possible location $l$ in the entire cortex, for every centroid, on every sweep of the Gibbs sampler. Even if we precache the geodesic distances between every pair of centroids (which would take dozens of gigabytes of memory) this operation is orders of magnitude more expensive than any other step in the parameter estimation.

We can reduce both the computational and memory demands of this step by only allowing the centroids to occupy a subset of the locations on the cortical surface. The cortical surfaces that we use have approximately 150,000 vertices in each hemisphere. We chose 5,000 random vertices in each hemisphere to serve as possible centroid locations. This immediately reduced both the computational and memory demands (of precaching the geodesic distances) by a factor of 30. Additionally, we exclude any potential centroids that are more than 150mm from the original centroid location. In practice these approximations result in slightly reduced fidelity of the reconstructed maps but a massive increase in performance. In the future we might be able to get around even these limitations by first doing a fast and coarse model fit and then following up by running a few iterations with less (or no) subsampling to fine-tune the parameters. 

\section{Running PrAGMATiC on an example dataset}
To test PrAGMATiC we applied it to a motor localizer dataset. This dataset consisted of BOLD fMRI data collected from 12 subjects while they underwent a 10-minute motor localizer. During the scan subjects were occasionally prompted to wiggle their fingers, wiggle their toes, move their mouths, generate internal speech, or generate frequent saccades. Each condition occurred in five 20-second blocks throughout the 10-minute scan. Standard fMRI preprocessing was applied, and then a linear model was used to find the change in BOLD response of each voxel in each condition relative to the mean BOLD response. The linear model weights for each condition were then projected onto cortical surface models that were specially constructed for each subject. The five resulting maps (one each for the foot, hand, mouth, speak, and saccade conditions) were then used as input data for PrAGMATiC.

We fit a PrAGMATiC model with 384 centroids using data from the left cortical hemisphere of 11 of the 12 subjects. For this model we fit the subject-specific edge lengths, $\Lm$, and did not fit spring constants, $K$. The model fitting ran for 4000 iterations. On each iteration we generated $J=5$ samples from $P(H;L,K)$ for each subject. The target entropy for the visible unit distribution $P_V$ was set to 0.2 (i.e. 20\% of the maximum possible entropy). The target entropy for the centroid distribution $P_h$ was set to 0.33 at the first iteration and then slowly lowered to 0.2 over the course of learning. The standard deviation of the visible unit emission model, $\sigma_V$, was initially set to 16.0 and then was updated according to the entropy of the visible unit distribution. The inverse spring temperature, $\beta$, was initially set to 0.01 and then was updated according to the entropy of the centroid distribution. The learning rate for the spring lengths, $\epsilon_L$, was set to 20.0. The learning rate for the area functional means, $\epsilon_M$, was set to 0.005. The learning rate for the subject-specific spring lengths, $\epsilon_\Lm$, was set to 1.0. The standard deviation of the Gaussian prior on the subject-specific spring lengths, $\sigma_\Lm$, was set to 25.0.

This model used seven anatomical landmarks to ensure that the maps approximately matched across subjects. These landmarks were specified manually in each subjects as vertex locations on the cortical surface. These landmarks were the top and bottom of the central sulcus, the front (genu) and back (splenium) of the corpus callosum, the frontal pole, the occipital pole, and the temporal pole.

\begin{figure}[htb!]
  \begin{center}
    \includegraphics[width=\textwidth]{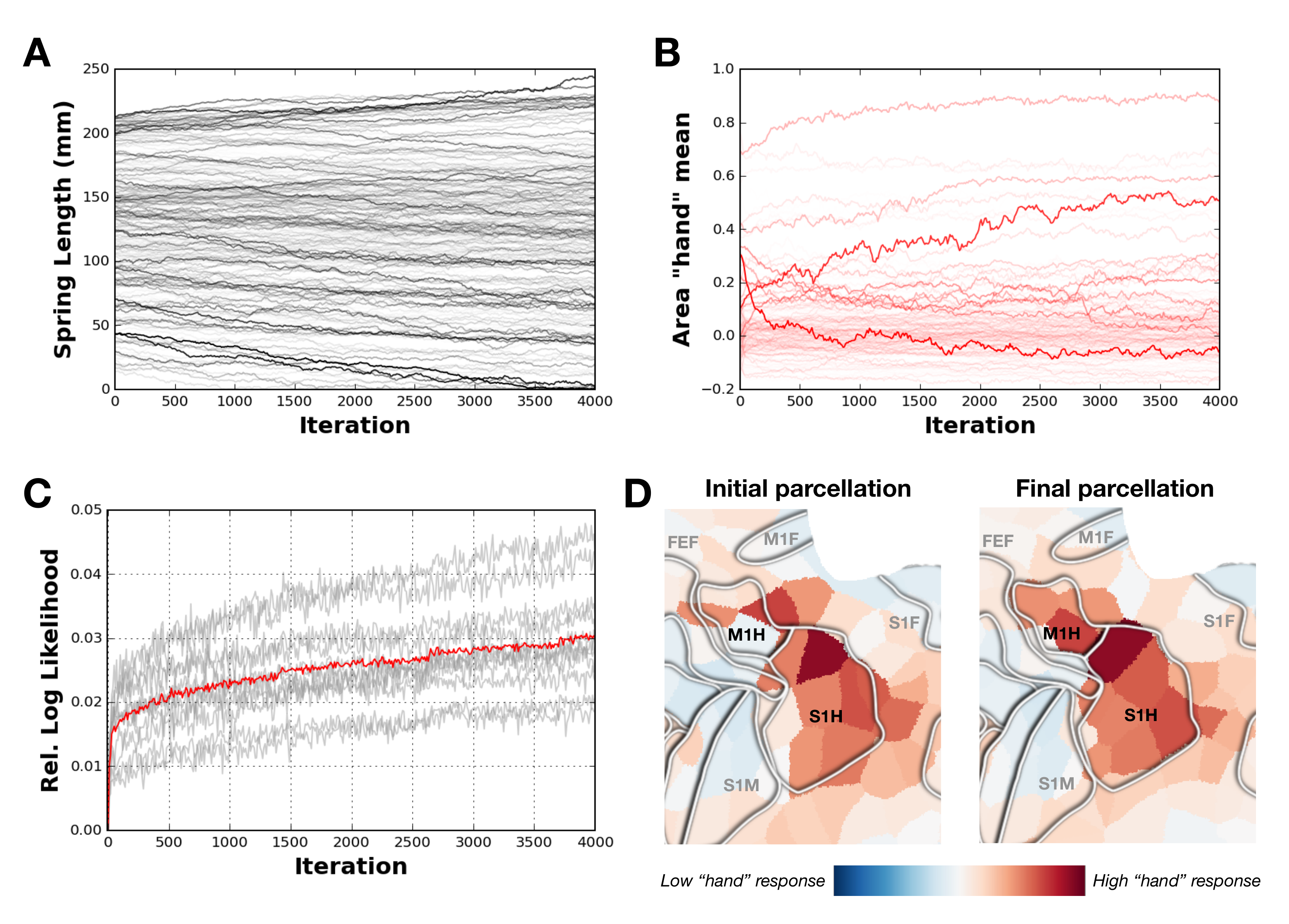}
  \end{center}
  \caption{\textbf{PrAGMATiC results on an example motor localizer dataset.} \textbf{(A)} Evolution of spring lengths over the course of learning for springs between one particular landmark (bottom of the central sulcus) and each area. Darker lines show spring lengths that changed more. \textbf{(B)} Evolution of area functional means over the course of learning for the functional dimension corresponding to hand movement. Darker lines show functional means that changed more. \textbf{(C)} Evolution of the mean log likelihood of the observed data given the model over the course of learning. Gray lines show individual subjects, red line shows the mean. Zero is defined as the likelihood at the first iteration. \textbf{(D)} Flattened cortical maps showing initial and final predicted hand response maps for a region around the central sulcus. Darker red means higher predicted hand response. White outlines show manually defined regions such as the primary motor hand area (M1H) and primary somatosensory hand area (S1H). The final map matches the manually defined areas much more closely than the initial map.}
\end{figure}

Some results from this model are shown in Figure 2. In Figure 2A we show how the spring lengths evolved for springs connecting one particular landmark (the bottom of the central sulcus) to all the centroids. The line for each spring is shaded according to its total change over the course of learning, so that darker lines correspond to springs that changed more. Here we see that many of the spring lengths changed relatively monotonically over the course of learning, while others moved both up and down. 

In Figure 2B we show how the area functional means evolved for one particular functional dimension corresponding to hand movement in the motor localizer. Similar to Figure 2A, the line for each area is shaded according to its total change over the course of learning. Here we see that a few areas seem to become much more selective over the course of learning, while many are unchanged. In contrast to the spring lengths where the changes appeared mostly monotonic, the area means seem to stabilize after a few thousand iterations.

In Figure 2C we show how the likelihoods of the observed data given the model evolved. The grey lines indicate individual subjects and the red line shows the mean. Likelihoods were computed at each iteration as the mean log likelihood across all the visible units given the model state. To ensure that likelihoods are comparable even though the visible unit standard deviation, $\sigma_V$, is changing during learning, we computed all likelihoods with $\sigma_V=1$. The likelihood plot for each subject is shown relative to their likelihood at iteration 0. Here we see that the likelihood increases monotonically throughout learning, confirming that the learning rules we derived are having their intended effect.

In Figure 2D we show predicted maps for the hand movement dimension in a region around the central sulcus at the first iteration and the final iteration. White outlines show the locations of manually defined regions including the known hand-responsive primary motor hand area (M1H) and primary somatosensory hand area (S1H). Here we see that hand-responsive areas in the initial map are not closely matched to the manual definitions. In the final map, however, boundaries of the highly hand-responsive areas closely match the manually drawn boundaries. This shows that PrAGMATiC is effectively learning the structure of the cortical map.

These results show that the PrAGMATiC learning rules derived here can be used to effectively learn a generative model of cortical maps. Future work on this dataset will be directed at comparing PrAGMATiC to other methods of cross-subject mapping.

\section{Future Directions}

Because PrAGMATiC is described as a hierarchical Bayesian network it is easy to extend in various ways. Here we describe some possible future extensions to PrAGMATiC that we believe will prove highly useful for neuroimaging research.

\paragraph{Extended hierarchical models.} One interesting future use case of PrAGMATiC will be comparing cortical maps between groups of subjects (e.g. patient groups vs. controls) or between different conditions (e.g. different attentional states) within the same subjects. This can be done by extending the PrAGMATiC hierarchical model to include group-specific area functional means, spring lengths, or spring constants. Standard Bayesian model comparison techniques could then be used to test specific hypotheses about the differences between cortical maps. For example, we could test whether two groups of subjects differ only in area functional means and not spring lengths, or vice versa. This type of cortical map comparison is not possible using any current techniques.

\paragraph{Generic map generating functions.} One obvious extension is using map generating functions other than the Voronoi tessellation detailed above. That is, in the energy function for the visible units, we can more generally write:
\begin{equation}
E(V|H;M) = \f{\sigma_V^{-2}}{2} \sum_{l,s} \tnorm{v_{ls} - f(H,M)_{ls}}^2
\end{equation}
where $f(H,M)$ is a function that takes an arrangement, $H$, and functional means, $M$, and converts them into a vector over all visible units in each subject. The only requirement that we have for this function is that it must be differentiable with respect to $M$ so that we can define:
\begin{equation}
\pderiv{E(V|H;M_i)}{M_i}{} = \sigma_V^{-2} \sum_{l,s} \left( v_{ls} -  f(H,M)_{ls}\right) \pderiv{f(H,M)_{ls}}{M_i}{}
\end{equation}
For example, this function could take the form of a smooth spline interpolation between the area centroids. Each possible map generating function could be thought of as a hypothesis about how cortical maps are organized. Then standard Bayesian model comparison techniques could be used to compare these hypotheses. This would enable us to test, for example, whether homogeneous functional areas (as used here) or smooth gradients better describe certain cortical maps.

\paragraph{Covariance-based models.} The model described here assumes that functional values are directly comparable across subjects. In resting state functional connectivity experiments there is no first-order relationship between functional values across subjects, but the covariance of functional responses across areas seems to be conserved \citep{Buckner2008}. PrAGMATiC could potentially be extended to capture these types of relationships by modeling the covariance matrix of visible unit responses. 

That is, we could assume that the covariance of $T$ measured functional values (e.g. a length $T$ resting state timecourse) across the $N_V$ total visible units within a subject is drawn from a Wishart distribution: $V^TV \sim \mathcal{W}_{N_V}(\Sigma, T)$. Here $\Sigma$ is an $N_V \times N_V$ symmetric matrix. Instead of learning parameters that describe the mean functional values within each area, $M$, we would learn parameters that describe the functional covariance between areas, $\Omega$, such that $\Omega_{ij}$ is the average covariance between visible units in area $i$ and visible units in area $j$. We define $\hat\Sigma_{kls}$, which is the model's best guess at the covariance between visible units $k$ and $l$ in subject $s$, as $\hat\Sigma_{kls} = \Omega_{c(H,k,s), c(H,l,s)}$. 

Then we could learn the covariance parameters, $\Omega$, the spring lengths, $L$, and the spring constants, $K$, so as to maximize the mean log likelihood of the observed covariances across subjects. This covariance-based formulation would allow us to learn PrAGMATiC parcellations using only resting state data.

\paragraph{Direct area-wise model fitting.} In early applications of PrAGMATiC we have assumed that the functional values for each visible unit capture functional selectivity. For example, these values might be derived from low dimensional projections of voxel-wise model weights \citep{Huth2012}. Yet it would also be possible to combine the voxel-wise model fitting step into the PrAGMATiC model, yielding an end-to-end Bayesian model. 

In voxel-wise modeling (VM) a complex natural stimulus is shown to subjects and then timecourses representing features of interest are extracted from the stimulus. VM fitting then uses ridge regression (or another regularized linear model) to estimate how the BOLD response in each voxel is influenced by each feature. Rather than estimating a PrAGMATiC model using the already estimated VM weights, we could directly estimate a separate linear model for each PrAGMATiC area. The functional timecourse for each visible unit, $v_{ls}(t)$ could be modeled as a multivariate Gaussian distribution: $v_{ls}(t) \sim \mathcal{N}(\mu_{c(H,l,s)}, \sigma^2_V I_T)$. Here we define $\mu_i$, the mean timecourse for area $i$, as a linear combination of the feature timecourses, $X$: $\mu_i = X \beta_i$. During model fitting we would learn the linear weights, $\beta$, the spring lengths, $L$, and the spring constants, $K$.

\paragraph{Automatic ROI labeling.} PrAGMATiC could also be used to automatically label well-known ROIs, such as the fusiform face area (FFA) \citep{Kanwisher1997} and parahippocampal place area (PPA) \citep{Epstein1998}. In this use case we would assume that the area membership of each visible unit is not known, but the functional values have been observed. For example, we might estimate a PrAGMATiC model using localizer data from a large number of subjects. Then we would assign each area a label, such as ``FFA'', based on anatomical location and functional selectivity (many areas may be assigned to each ROI). After acquiring localizer data for a new subject, we could then label known ROIs by applying the estimated and labeled PrAGMATiC model. That is, we could use Gibbs sampling and simulated annealing to optimize the arrangement of the areas in the new subject so that the probability of the observed data is maximized, finding $H^* = \mbox{argmax}_H P(V^{obs}_{new} | H; L, K, M)$. Then we would define FFA in the new subject as the collection of visible units belonging to the areas that were labeled as FFA. This procedure would greatly simplify the process of labeling known ROIs, which is currently laborious and manual.

\paragraph{Representational similarity.} One method for analyzing fMRI data that has gained popularity in recent years is representational similarity analysis (RSA) \citep{Kriegeskorte2008}. In this method an ROI is selected (or an anatomical searchlight is used) and then the covariance of responses within that ROI are computed across stimuli to produce a representational dissimilarity matrix (RDM) for the ROI. These RDMs can then be compared across ROIs or between ROIs and models. One issue with this method is that it either requires prior definition of an ROI across which the RDM is to be computed, or uses an anatomical searchlight. Instead, PrAGMATiC could be used to find areas that have similar representational dissimilarity and anatomical location across subjects.

In this method the functional means would be replaced with mean RDMs for each area. Then on each iteration we would compute an RDM for each area, and evaluate the likelihood of the individual RDM given the mean RDM. This could use a similar Wishart formulation to the covariance-based model above.

\bibliography{library}{}

\end{document}